Running Head: VOCABULARY AND THE BRAIN

Vocabulary and the Brain: Evidence from Neuroimaging Studies

Tom A. F. Anderson

Graduate Institute of Network Learning Technology

Prof. Ruan, C.-H.

Institute of Cognitive Neuroscience

National Central University

# Abstract

In summary of the research findings presented in this paper, various brain regions are correlated with vocabulary and vocabulary acquisition. Semantic associations for vocabulary seem to be located near brain areas that vary according to the type of vocabulary, e.g. ventral temporal regions important for words for things that can be seen. Semantic processing is believed to be strongly associated with the ANG. Phonological ability has been closely related to the anterior surfaces of the SMG. Pathways through the posterior SMG are thought to link the anterior SMG and the ANG. In vocabulary tasks, mediotemporal structures may be related to long-term memory processing, with left hippocampal and parahippocampal regions related to long-term and working memory, respectively. Precentral structures are associated with phonological retrieval. Furthermore, many more regions of the brain are of interest in vocabulary tasks, particularly in areas important for visual and auditory processing. Furthermore, differences between brain anatomies can be attributed to vocabulary demands of different languages.

# Introduction

The human brain spontaneously reacts to words. Electrical recordings of the scalp show activity 160 milliseconds after exposure to a word in a task of giving the use of a word (Posner & Raichle, 1997). Furthermore, different activations of different areas are observed for different vocabulary tasks, enabling the production of hypotheses about which areas of the brain regard vocabulary and why. According to Blakemore and Frith (2005), humans begin the process of life-long vocabulary learning as babies, fastmapping words to objects, learning 20-50 words by 18 months of age. Between 18 – 24 months of age, the rate of vocabulary acquisition increases dramatically, and five-year-old children generally have a 2000 word vocabulary in their native language. Adults also learn vocabulary at a fast rate. Nonetheless, the neurological mechanisms underlying vocabulary processes are still under investigation, so there is a lot that can be learned from neuroscience research. Therefore, vocabulary and vocabulary learning is an important part of the learning sciences that merits a literature survey of some of the related neuroscience such as this paper.

Language in general, and word learning specifically, are commonly researched topics in experimental psychology and cognitive science. In experimental psychology, the Stroop effect demonstrates that the reaction time of word naming varies according to other factors. Around 1973, autoradiography of living human subjects came about (Posner & Raichle, 1997), allowing investigations into changes that occur in the brain. By pairing cognitive science tasks with neuroimaging techniques, such as pairing the Stroop effect with positron emissions tomography, we are able to come to understand more closely the interworking of the brain.

Some areas of the brain have long been known to be strongly related to language ability. For example, Broca's area, a portion of the left inferior frontal gyrus, has long been associated with learning one's native language. Using fMRI, Musso et al. (2003) showed that when learning real grammar of a second language, activation in Broca's area increases over time (See Fig. 1). Such research results and innate human interest in language, especially in word and vocabulary learning, have spurred neuroscientists to use modern neuroimaging techniques for the investigation of word recognition and of word learning. Techniques such as ERP, PET, and fMRI have been used for investigations of areas of responsibility for vocabulary in the brain. These types of tests can often provide more insight into the correlational basis obtained from analysis of patients with brain lesions. And they can show how many areas of the brain relate to language. In this paper, two terms in particular occur quite often; therefore, the supramarginal gyrus will be abbreviated as SMG, and the angular gyrus will be abbreviated ANG.

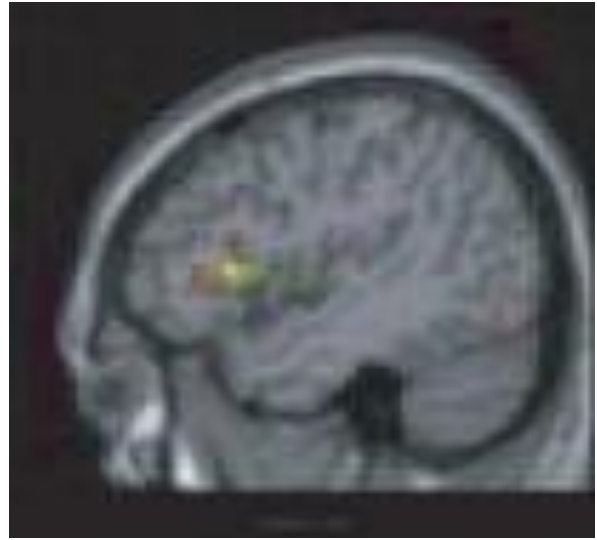

**Figure 1. Activation in learning grammar rules of real Italian (yellow) and Japanese (red) indicate increasing activation in Broca's area.**

## Common Neurological techniques

Humans have been intensely interested in studying the brain since before the time of Aristotle and Descartes, famous scholars whose conjectures about the workings of the brain influenced many scholars. More modern conceptions of the mind from luminaries such as Sechenov, Scherrington, Hebb, Konorsky and Luria brought understanding of the biologic nature of the mind (Posner & Raichle, 1997). Beginning with studies of lesions, and progressing to computer-aided neuroimaging techniques, the following is a brief summary of basic techniques used in research cited in this paper.

*Positron Emissions Tomography (PET)* – Isotopes (e.g. carbon, nitrogen, oxygen, and fluorine) emit gamma radiation that rapidly decays. Emissions from these elements located in an area of the body such as the head can be detected to produce tomographs of the brain, reconstructed computer images that are most usually in color (Posner & Raichle, 1997).

*Functional Magnetic Resonance Imaging (fMRI)* – According to Posner and Raichle (1997), by examining the differences in the proton signals, functional magnetic resonance imaging, or fMRI, allows researchers to investigate the functioning of the brain. In fMRI, magnetic fields orient the atoms that are to be observed. Subsequent application of radio wave pulses induce radio signals that inform as to the number of atoms as well as the chemical environment, about the anatomy and function of the subject. MRI is now often referred to as fMRI. Using fMRI, changes in brain blood oxygen can be measurable at resolution of about 1-2 millimeters, indicating neural activity (Ogawa et al., 1990, as cited by Binder, et al., 1996). A significant advantage of fMRI is that it does not induce changes in brain tissue, as do PET and x-ray tomography.

*Voxel-based morphometry (VBM)* – A whole-brain technique, VBM uses structural magnetic resonance images (Green, 2007). First, brains are normalized to eliminate individual variations in brain shapes. Then, once the positions and sizes of the gyri of each of the brains to be studied are correlated, the technique is used to determine small-scale differences in the concentration and volume of grey matter and white matter.

*Brain Lesions* – Lesions are areas of tissue damage. When they occur in the brain, they may cause neural structures to function suboptimally. Researchers have learned a great deal about from associating lesions with inabilities of patients. For language in particular, double dissociation is often seen in patients with lesions in one area and not in another.

### *Double Dissociation*

Language as cognitive processes is linked to neural mechanisms, many of which are distinct from other brain systems. For evidence of this, we can see patients with brain lesions who have disorders of language, yet no other discernable problems; also, there are other patients with no language problems, yet many other problems. We call this "double dissociation," and it is a general indication of modular, independent systems. Even within closely related tasks that fall under the language umbrella there is double dissociation: for example, word naming and picture naming are double dissociated. Also speaking and singing are areas of double dissociation because: there are some subjects with lesions in one area who are able to speak yet cannot sing; there are other subjects with lesions in another area who can sing, yet cannot speak. Double dissociation gives strong indication that two areas of the brain are dedicated to different processes.

# Research findings

A brief introduction to language areas in the brain will be reported, followed by findings more specific to vocabulary.

### *Language areas of the brain*

When subjects are exposed to stimulus such as language, subsequent activation in the brain can be observed. By comparing the activation that results from language to activation that results from meaningless sounds, it can be inferred, through the subtractive method, that the remaining areas of activation are specific to language. Binder et al. (1997) performed such an imaging experiment and found results that both confirmed and went against classical models.

Binder et al. found that language processing primarily showed activation in the left cerebral hemisphere involving a network of regions in frontal, temporal, and parietal lobes. They found several things not in concordance with traditional language areas:

> "(1) The existence of left hemisphere temporoparietal language areas outside the traditional "Wernicke's area," namely, in the middle temporal, inferior temporal, fusiform, and angular gyri;
>
> (2) Extensive left prefrontal language areas outside the classical 'Broca's area'; and
>
> (3) Clear participation of these left frontal areas in a task emphasizing "receptive" language functions." (Binder et al., 1997).

Activation for a semantic decision test was observed in four distinct cortical language-related areas: (1) Researchers observed activated cortex on both sides of the superior temporal sulcus. The middle temporal gyrus in the left hemisphere was mostly activated. Additionally, activation was also evident in the inferior temporal gyrus, fusiform, and parahippocampal gyri in the ventral temporal lobe. They noted that some areas were more strongly activated in response to meaningless audio stimulus than to the stimulus in the semantic decision task.

(2) Secondly, in the middle frontal gyrus, the rostral and caudal areas were active while the midportion was not. Activation occurred in a great deal of the superior frontal gyrus anterior to the vertical AC line. Left medial frontal activation spread ventrally to involve a portion of the anterior cingulate gyrus. Smaller activation was also noted in the anterior cingulate and superior frontal gyrus in the right hemisphere.

The anterior cingulate system is seen active in experiments into the Stroop effect (Posner & Raichle, 1997) perhaps due to its function in the inhibition of the automatic response to a word when an ink-color must be reported.

(3) Activation was observed in the ANG, which would correspond to phonological processing, as cited by Lee et al. (2007).

(4) A perisplenial region including the posterior cingulate, ventromedial precuneus, and cingulate isthmus, clearly distinct from the activation observed in the meaningless sound stimulus condition.

Another large region of activation in this task was the right posterior cerebellum. The results of the brain imaging are shown in Figure 2.

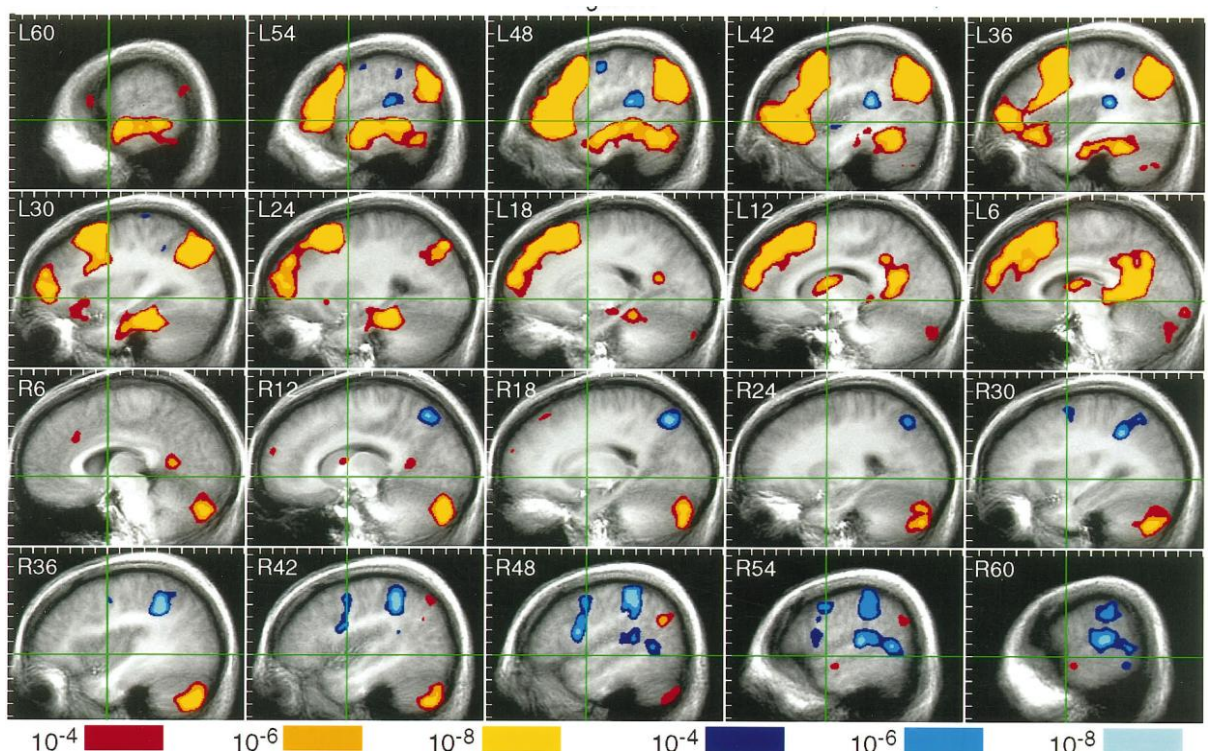

**Figure 2. Activation of the brain in semantic processing of hearing spoken language obtained by subtracting activation caused by non-language audio stimuli. From Binder et al., 1997.**

Using PET scans of the angular cingulate gyrus and subtracting for sensory and motor activations lead Posner and Raichle (1997) to concur that this area is responsible for attention. Activation of the anterior regions of the brain was also detected in the basal ganglia, corresponding to the detection of colors, motions and forms. They suggest that frontal areas generally demonstrate activity when active processing is demanded by the experimental task.

### *Word recognition*

As cited by Nobre & Plunkett (1997), two regions specialized for word recognition—both on the ventral surface of the temporal lobes—have been identified. This area of the temporal lobes is strongly associated with object recognition in both humans and primates. Using intracranial electrodes, it was determined that there was a focal region in the posterior ventral extrastriate visual cortex that responded to strings of letters and to words, but did not react to other meaningful visual stimuli, nor to orthography, phonology or semantics. Additionally, another more anteriorly located region was active for letter strings that obeyed orthographic and phonological rules, with varied responses given based on content and context. Face recognition regions were located near each of these areas of word recognition, which would indicate that face recognition and word recognition share parallel organization. It appears that word and object recognition are closely related within the human brain.

Typical hypotheses for language in the adult brain are that the left hemisphere is dominant, the left-temporal lobe is involved in language comprehension, and the left-frontal lobe is involved in expressive language functions. However, some things are clear from children with focal brain injuries. For children with focal brain injury, we can see that the two hemispheres are equally able to support language. Children less than 6 months of age with brain injury to only one hemisphere did experience language impairments, though by age 5 or 6 no language impairments are detectable. At around 18 to 24 months of age, an explosive growth in vocabulary is typically evidenced; however, this growth is disrupted in those of 19-31 months of age with left temporal lesions. This implies that for language, brain localization is organized according to the experience of the individual, rather than language centers being simply assembled (Nobre & Plunkett, 1997).

## *Face, animal and tool naming*

The naming of different things—such as faces, tools, and animals—have been shown to activate specific regions, especially in the temporal pole. Patients may have deficits in tool naming and not animal naming, or deficits in animal naming and not tool naming. However, those with naming deficits for both face and tool naming always also have naming deficits for animals, as the region of the brain for animals is located between the other two. The location and separation of these areas of activation determined by lesion have been confirmed by PET studies (Damasio et al. 1990, cited by Nobre and Plunkett, 1997). Neuroimaging suggests that the representations of attributes of objects are located near the cortical regions that mediate perception of those attributes. Therefore, the coding of vocabulary occurs throughout widespread regions that code sensory, motor, or functional attributes, and additionally in regions that integrate or bind together such coding of sensory (visual, auditory, etc.) motor, or functional attributes.

Lesions in the left hemisphere and specific non-perisylvian regions of the temporal lobe were correlated with naming deficits that correlated with the location of the lesions (Martin, et al., as cited by Nobre & Plunkett, 1997). With lesions in the temporal lobe came deficits in naming faces. More posterior ventral temporal lesions extending to lateral sites along junction of temporal-occipital-parietal cortices lead to impaired naming of tools. The ventral temporal regions have been linked to the naming of animals (a task that primarily involves visual form, reliant on ventral pathways of visual system). The left lateral middle temporal gyrus and left premotor regions is linked to naming tools (reliant on motor/visual motion associated with the tool, reliant on the dorsal visual pathway). Finally, the extrastriate visual cortex, ventral temporal lobe, and premotor regions are all activated in naming both animals and tools.

As observed by PET, brain area activations were highly similar for both pictures and words. A posterior lateral region over the middle and superior temporal gyrus was activated for generating verbs. Lesions in the left premotor and prefrontal region produce impairment to the naming of actions (verbs). Additionally, left premotor areas in or near Broca's area are activated for generating words for actions (Martin et al., 1995, as cited by Nobre & Plunkett, 1997). We might infer that these motor-related regions are linked to naming actions because actions require motor activity.

### Concrete nouns and function words

Neuroimaging shows highly non-overlapping brain activation for concrete nouns and function words (Nobre, et al., 1997, as cited by Nobre & Plunkett, 1997). This may be due to brain region specificity for different functions.

Left-hemisphere brain regions linked with visual and associative functions in semantic tasks are activated by concrete nouns, which have strong visual and mnemonic associations. Areas of activation include the inferior frontal cortex, posterior middle temporal cortex, medial temporal lobe regions including the hippocampus and surrounding cortex, lateral anterior temporal cortex, posterior cingulate cortex and superior dorsolateral prefrontal cortex.

Left-hemisphere brain regions linked to phonology and grammar are activated by function words, which are used for linking and sequencing other items. Areas of activation include the superior temporal sulcus and middle temporal gyrus, SMG at the end of the sylvian fissure, precentral sulcus in premotor cortex in the region of Broca's area, and inferior frontal gyrus and precentral gyrus in motor cortex.

### Covert naming

Comparing the viewing of objects with the passive viewing of non-objects, researchers using functional magnetic resonance imaging (fMRI) were able to observe that silently naming objects activates certain areas of the brain more than others. Subjects were trained to use particular names for pictures of objects and non-objects, and subsequently—in the experiment—subjects internally named the objects to themselves. The baseline was determined as the brain activation occurring for observation of non-objects that could not be named as in the experimental condition, for which words were known. It was observed that in the experimental condition, both left and right hemispheres were activated in the frontal, parietal, and mediotemporal lobes, as seen in Figure 3. Such bilateral occipital activation might be observed as a result of visual processing (Menard et al., 1996 as cited by Ellis et al., 2006),

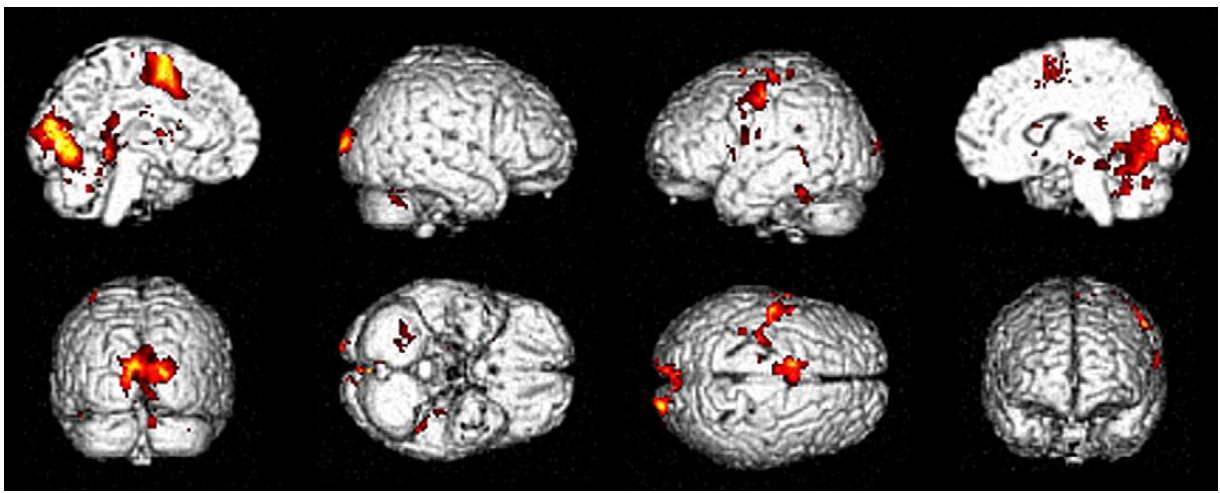

**Figure 3. Areas of brain activated in covert naming task but not in baseline task. From Ellis et al., 2006.**
perhaps by triggering the imagination of visual features that are not seen in an object because they are obscured by another part of the same object. In non-objects, this kind of stimulation is not observed, because subjects do not have experience about what visual features may be obscured.

In the left hemisphere: activations in precentral and medial frontal gyri; postcentral, SMG and ANG in the parietal; and the middle temporal gyrus, the parahippocampal, lingual, and fusiform gyri in the temporo-occipital ventral regions; and in the hippocampus. Activation of the precentral structures has been linked to phonological retrieval (Murtha et al., 1999 as cited by Ellis et al., 2006), which implies that the subjects were activating phonological structures though they were not speaking the name of objects aloud. The left hippocampal and parahippocampal regions may reflect use of long-term and working memory in this task. These results are consistent with naming objects as measured in functional neuroimaging studies (Murtha et al. 1999; Price, 2000; Humphreys and Price, 2001; as cited by Ellis et al., 2006). Additionally, mediotemporal structures may also be related to long-term memory retrieval (Moscovitch et al., 2006, cited by Ellis et al., 2006). The left parieto-temporal junction activation shows that in this task, working memory might be involved in retrieving and checking the nonverbalized names.

Further areas of activation include, in the right hemisphere: Insula, postcentral gyrus, parietal lobe and the parahippocampal gyrus in the ventro-temporal region. Furthermore, activation was seen in the body of the left caudate, the left claustrum, and bilaterally in the putamen.

### *Age of acquisition effect in covert naming*

Some words are learned much earlier in life than others (e.g. the word "brain" is learned years before the word "transcranial"). In word naming tasks, subjects are faster to respond and name words that are learned earlier in life. Brain activation varies according to when a word is learned, depending on whether the subject learned the word early in life or not. Activation in posterior parts of the middle gyri in the occipital poles is evident for early acquired words; left middle occipital and fusiform gyri showed greater activation for those words that are acquired later. Neuroimages for the two conditions may be found in Ellis et al. (2006). Figure 3 in this work shows neural activation for a more general condition.

*Vocabulary Acquisition in Adolescents*

Gray matter density has been shown to increase as a result of learning, even in adults, and measures of skill have been correlated with the changes. Through functional imaging, it has been determined that in the inferior parietal lobe (IPL), two areas of activation during vocabulary acquisition (word learning) are located in the vicinity of the anterior and posterior surfaces of the SMG (Breitensten et al., 2005 and Cornelissen et al., 2004, cited by Lee, et al., 2007). Proficiency at a second language can be predicted by gray matter density in an area of the inferior parietal lobe—the posterior SMG (Mechelli, 2004, cited by Lee, et al., 2007)—though the posterior SMG is not typically activated during functional imaging studies of word processing (cited by Lee, et al., 2007). The posterior surface of the SMG is located between the anterior SMG and the anterior ANG. The anterior surfaces of the SMG have been associated with phonological abilities, such as the ability to pronounce novel words (Cornelissen et al., 2004, cited by Lee et al., 2007). The anterior ANG region has been associated with semantic processing (as cited by Lee, 2007).

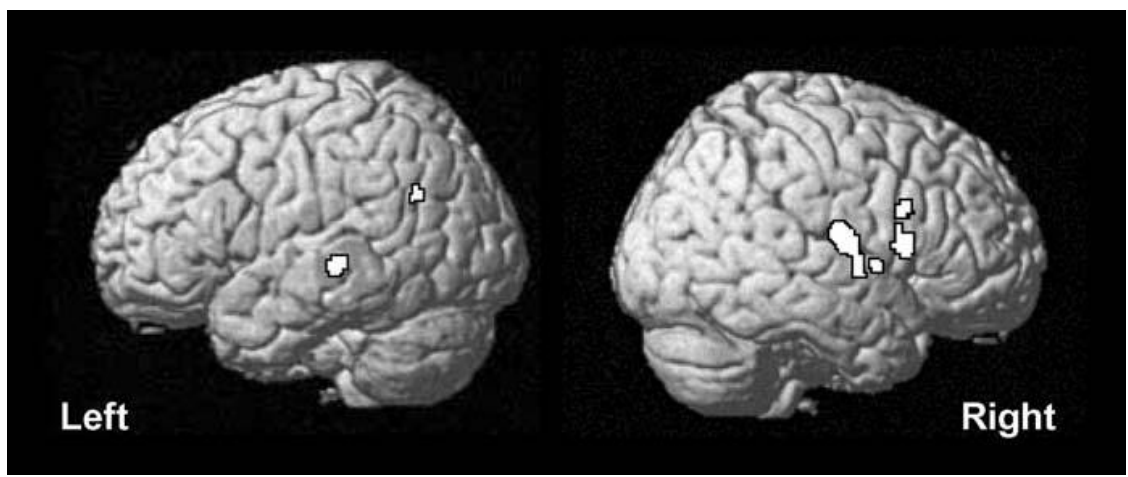

**Figure 4. Areas of increased grey matter density in speakers of Chinese as compared to speakers of English, after removing density variations attributable to ethnicity. From Lee et al, 2007.**

Lee et al. (2007) used voxel-based morphometry (VBM) to investigate the effects of vocabulary acquisition on the gray matter density between the anterior and posterior SMG.

The researchers performed brain image analysis to compare 32 adolescents with similar verbal IQs to find differences that could be attributed to word knowledge alone. A separate measure of verbal fluency, or verbal IQ, showed that gray matter density is correlated with vocabulary knowledge even for individuals if they have similar verbal IQs. Therefore, they were able to show strong correlation between gray matter and verbal knowledge. They reported that gray matter density in the posterior SMG is most strongly correlated with vocabulary knowledge among adolescents.

The results of monkey studies show that between similar regions in the macaque brain, pathways directly linking the two regions only go through the region that corresponds to the posterior SMG. Lee et al. also conducted human studies, showing through tractographic analysis of MRI data that it appears that there is only one pathway from the anterior SMG to the ANG, which goes through the posterior SMG. It is then hypothesized that the role of the posterior SMG may be crucial in vocabulary learning, as the posterior SMG links areas that are independently responsible for phonological and semantic processes, respectively.

One thing that I consider interesting that was mentioned in Lee's study is that monkey brains have an analogue to the posterior SMG pathway from ANG to anterior SMG. This demonstrates that vocabulary abilities are related to regions of the brain that have not evolved for language alone.

### *Brain differences between speakers of Chinese and English*

Areas of the brains of Chinese people have been shown to be different than in Caucasian brains, with increased gray matter density in the right parietal, left frontal and left temporal regions. Caucasian brains, on the other hand, show increased density in left superior parietal regions (Kochunov et al., 2003, as cited by Green, et al., 2007). In order to probe these differences, Green, et al. (2007) used VBM to investigate differences between speakers of Chinese and speakers of English as a first language. The researchers controlled for underlying differences between Chinese and European brains, and investigated non-Asian speakers of Chinese as well as Chinese people for whom. They were able to confirm that the effect observed was due to language, not ethnicity.

In two areas of the left hemisphere (the middle temporal gyrus and in the superior temporal gyrus, anterior to Heschle's gyri) and in two regions of the right hemisphere (the superior temporal gyrus anterior to Heschle's and a region in the inferior frontal gyrus), speakers of Chinese show highly significant enhanced gray matter, as seen in Figure 4. Increased gray matter density was observed in the posterior SMG region in speakers of more than one language, no matter Asian or Western. That area corresponds to the linking of semantics and phonology seen in the previous adolescent vocabulary acquisition study (Lee, et al., 2007). Green et al. conclude that the increase in density in gray matter in these areas may be due to the acquisition of second language vocabulary.

## Discussion

Though the studies that I have outlined above do illustrate that certain areas are activated during word recall, it cannot be concluded that for language learning, particularly for word learning, that there is therefore a system dedicated for that purpose. In fact, it has been shown that there is no dedicated system for word learning, as shown by a study which children learn the meaning of a new word after being exposed to it only a few times (Markson & Bloom, 1997). In contrast, many other aspects of language learning have been associated with dedicated systems (e.g. Broca's area for grammar learning).

In future research, investigators will be able to do more long-term studies. In addition to learning about neuroanatomy, we will be able to find how that anatomy became that way. As discussed by Green et al. (1997), it is unlikely that the differences in gray matter density allow individuals to acquire more vocabulary; rather, it is likely that gray matter density reflects the binding of sound and meaning, thus gray matter likely grows as a result of learning. For an example of a vocabulary related task that could be undertaken in this manner, I predict that researchers may be able to determine the interworking of the spacing effect, one of the earliest tenets of experimental psychology, which often involve the training for word pairs, one in the subject's native language and the other in an unknown language.

Another interesting topic is that of comparison of the brain and learning between Western and Asian languages. Nisbett (2003) pointed out that many cultural differences that exist between Westerners and Asians are related to differences between the languages, and that even the way that the languages are taught differs greatly, with Western languages focusing much more on learning nouns than verbs, and almost the opposite occurring in Asian languages. Green et al.'s (2007) study showed that speakers and learners of Chinese have different brain structures than those who do not know Chinese, largely due to phonological differences. There are many other comparative studies that can show more clearly other differences in brain structure due to vocabulary differences.

Finally, foreign language experts such as Steven Krashen sometimes criticize neuroscientists as it places too much attention to focus on meaningless input (Krashen, 2008). In all the studies we see regarding vocabulary, including those reported above, attributed to language in general, yet language is actually more about comprehension than about words or phonology. We saw activation in many areas of the brain in the study by Binder et al. (1997), meaning that the task of analysis in such comprehension-related neuroscience experiments would be great. Perhaps as supercomputers become more affordable, neurolinguistic studies can be conducted for comprehension.

## Conclusion

After introducing the topic of word learning, I summarized major neuroscientific techniques, especially neuroimaging. I presented research findings from a number of brain imaging studies. In the discussion, I drew out some conclusions that could be made on the basis of these studies, and also presented some studies that caution against overreaching these conclusions.

What I have learned from writing this paper is about which areas of the brain are most associated with word learning. I also learned that many of the relevant areas are located next to the sensory and motor associated with the word. Practically speaking, this implies that to promote word learning, we should promote rich learning that allows the learner to relate to the word with sensory (e.g. visual, phonological) and motor areas (e.g. movement). Additionally, the research provides an insight that the brain changes as we learn more vocabulary, no matter the age, as vocabulary is learned at all ages.